\documentclass [10pt]{article}
\usepackage[left=2cm,top=2.50cm,right=2cm,bottom=2.50cm]{geometry}
\usepackage{mathrsfs}
\usepackage{physics}
\newcommand*\rfrac[2]{{}^{#1}\!/_{#2}}

\usepackage{amsmath,amssymb,latexsym,color,cancel,graphicx,bbm,colortbl}
\usepackage[english]{babel}
\usepackage[latin1]{inputenc}
\usepackage{ragged2e}
\usepackage{cite}
\usepackage{subfig}

\begin{document}
\date{}

\title{$SU(1,1)\times SU(2)$ approach and the Mandel parameter to the Hamiltonian of two oscillators with weak coupling}
\author{J. C. Vega$^{a}$, D. Ojeda-Guill\'en$^{b}$\footnote{{\it E-mail address:} dojedag@ipn.mx}, and R. D. Mota$^{c}$} \maketitle

\begin{minipage}{0.9\textwidth}
\small $^{a}$ Escuela Superior de F{\'i}sica y Matem\'aticas, Instituto Polit\'ecnico Nacional,
Ed. 9, U.P. Adolfo L\'opez Mateos, Alc. Gustavo A. Madero, C.P. 07738 Ciudad de M\'exico, Mexico.\\

\small $^{b}$ Escuela Superior de C\'omputo, Instituto Polit\'ecnico Nacional,
Av. Juan de Dios B\'atiz esq. Av. Miguel Oth\'on de Mendiz\'abal, Col. Lindavista,
Alc. Gustavo A. Madero, C.P. 07738, Ciudad de M\'exico, Mexico.\\

\small $^{c}$ Escuela Superior de Ingenier{\'i}a Mec\'anica y El\'ectrica, Unidad Culhuac\'an,
Instituto Polit\'ecnico Nacional, Av. Santa Ana No. 1000, Col. San
Francisco Culhuac\'an, Alc. Coyoac\'an, C.P. 04430, Ciudad de M\'exico, Mexico.\\
\end{minipage}

\section*{Abstract}
We study the Hamiltonian of two isotropic oscillators with weak coupling from an algebraic approach. We write the Hamiltonian of this problem in terms of the boson generators of the $SU(1,1)$ and $SU(2)$ groups. This allows us to apply two tilting transformations based on both group similarity transformations to obtain its energy spectrum and eigenfunctions. Then, we obtain the Mandel $Q$-parameter and the second-order correlation function $g^2(0)$ of the photon numbers $n_a$ and $n_b$. It is important to note that in our procedure we consider the case of weak coupling.

\section{Introduction}

The simple harmonic oscillator has been studied for 300 years and still remains a fundamental piece in the study of the universe. It is undoubtedly one of the fundamental problems in all branches of physics, and in particular in quantum mechanics \cite{Bloch}. However, harmonic oscillators are not disconnected from each other, so it is important to study coupled harmonic oscillators.

One of the main applications of coupled harmonic oscillators was found in nonlinear physics, and a complete theory was developed in this field \cite{Fano,Schweber,Estes,Fetter,Kim,Han,Iachello,Han2,Jakub}. In Ref. \cite{Han} the authors study the Hamiltonian of two oscillators with position-position coupling and obtained the eigenfunctions of the Schr\"odinger stationary equation. Moreover, the study of quantum entanglement of these systems is of vital importance \cite{Joshi,Paz,Galve,Fillaux}, since interesting applications can be found from biophysics to quantum teleportation, as well as quantum cryptography, quantum code and quantum algorithms \cite{Romero,Fuller,Halpin,Ekert,Bennett,Shor,Samuel,Makarov}.

The aim of this work is to study the Hamiltonian of two oscillators with weak coupling from an algebraic approach. We obtain its eigenfunctions and energy spectrum by using similarity transformations in terms of the $su(1,1)$ and $su(2)$ Lie algebras and compute the Mandel parameter of this problem for the photon numbers $n_a$ and $n_b$.

This work is organized as it follows. In Sec. $2$ we introduce the Hamiltonian of two coupled oscillators in terms of the bosonic operators $a$ and $b$. Then, we introduce Schwinger realizations of the $su(1,1)$ and $su(2)$ Lie algebras to apply two tilting transformations in terms of these realizations to the Hamiltonian to diagonalize it. Assuming that the coupling is weak ($\lambda^2 \ll 1$), we can obtain the energy spectrum and the eigenfunctions of our problem. The Sec. $3$ is dedicated to compute the Mandel $Q$-parameter and the second-order correlation function $g^2(0)$ of our Hamiltonian for the photon numbers $n_a$ and $n_b$. Finally, we give some concluding remarks.

\section{The Hamiltonian of two oscillators with weak coupling}

The Hamiltonian which describes the behavior of a system of two coupled harmonic oscillators with identical free frequencies with position-position coupling is defined as \cite{Estes}
\begin{equation}
    H_{xy} = \frac{1}{2m} \left( p_x^2 + p_y^2 \right) + \frac{m \omega^2}{2} \left( x^2 + y^2 \right) + 2\kappa m \omega xy,
\end{equation}
where $\kappa$ is a constant that describes the strength of the coupling. This Hamiltonian can be written in terms of the bosonic operators as it follows
\begin{equation}
    H_{xy} = \hbar \omega \left( a^{\dag}a + \frac{1}{2} \right) + \hbar \omega \left( b^{\dag}b + \frac{1}{2} \right) + \hbar \kappa (a^{\dag} + a) (b^{\dag} + b).
\end{equation}

Inspired by this Hamiltonian, the purpose of this work is to study from now on the following more general Hamiltonian (with $\hbar = 1$)
\begin{equation}
    H = \omega (a^{\dagger} a + b^{\dagger} b + 1) + \lambda e^{-i \psi} (a^{\dagger} b^{\dagger} + a^{\dagger} b) + \lambda e^{i \psi} (b^{\dagger} a + b a). \label{Hamiltonian}
\end{equation}
Using the Schwinger realizations of the $SU(2)$ and $SU(1,1)$ groups with two bosons of Eqs. (\ref{su2}) and (\ref{su11ab}), we can transform this Hamiltonian to
\begin{equation}
    H = 2 \omega K_0 + \lambda e^{-i \psi} (K_+ + J_+) + \lambda e^{i \psi} (K_- + J_-).
\end{equation}
This Hamiltonian can be diagonalized if we apply a tilting transformation to the stationary Schr\"odinger equation $H\Psi=E\Psi$ as it follows \cite{Gerryberry,Nos1,Nos2,Nos3,Nos4}
\begin{equation}
D^{\dagger}(\xi)HD(\xi)D^{\dagger}(\xi)\Psi=ED^{\dagger}(\xi)\Psi.
\end{equation}
Here, $D(\xi)$ is the $SU(1,1)$ displacement operator defined as $D(\xi) = \exp(\xi K_+ - \xi^{*} K_-)$, with $\xi = - \rfrac{1}{2} \tau e^{-i \phi_\xi}$ (see Appendix A). Thus,
if we define the tilted Hamiltonian $H'$ and its eigenfunction $\Psi'$ as $H'=D^{\dagger}(\xi)HD(\xi)$ and $\Psi'=D^{\dagger}(\xi)\Psi$, from the similarity transformations (\ref{st1}) and (\ref{st2}) of Appendix B we obtain
\begin{eqnarray}
H' &=& 2\omega \left[ (2 \beta_\xi + 1) K_0 + \frac{\alpha_\xi \xi}{2 \abs{\xi}} K_+ + \frac{\alpha_\xi \xi^{*}}{2 \abs{\xi}} K_- \right] \nonumber\\
        &&+ \lambda e^{-i \psi} \left[ \left[ \frac{\xi^*}{\abs{\xi}} \alpha_\xi K_0 + \beta_\xi \left( K_+ + \frac{\xi^*}{\xi} K_- \right) + K_+ \right] + \left[ \frac{\xi^*}{\abs{\xi}} \alpha_\xi K_{-}^{(b)} + \frac{\xi}{\abs{\xi}} \alpha_\xi K_{+}^{(a)} + (2 \beta_\xi + 1) J_+ \right] \right] \nonumber\\
        &&+ \lambda e^{i \psi} \left[ \left[ \frac{\xi}{\abs{\xi}} \alpha_\xi K_0 + \beta_\xi \left( K_- + \frac{\xi}{\xi^*} K_+ \right) + K_- \right] + \left[ \frac{\xi^*}{\abs{\xi}} \alpha_\xi K_{-}^{(a)} + \frac{\xi}{\abs{\xi}} \alpha_\xi K_{+}^{(b)} + (2 \beta_\xi + 1) J_- \right] \right],
\end{eqnarray}
where $\alpha_\xi = \sinh{(2\abs{\xi})}$ and $\beta_\xi = \rfrac{1}{2} [ \cosh{(2\abs{\xi})} - 1]$. In this expression $\{K_{\pm}^{a},K_{\pm}^{b}\}$ are the one-boson ladder operators for the $SU(1,1)$ realization (see Eq. (\ref{su11a} of Appendix A). By grouping terms we can express the tilted Hamiltonian as
\begin{eqnarray}
        H' &=& \left[ 2 \omega (2 \beta_\xi + 1) + \frac{\xi^*}{\abs{\xi}} \alpha_\xi \lambda e^{-i \psi} + \frac{\xi}{\abs{\xi}} \alpha_\xi \lambda e^{i \psi} \right] K_0 \nonumber\\
        &&+ \left[ \frac{\omega \alpha_\xi \xi}{\abs{\xi}} + (\beta_\xi + 1) \lambda e^{-i \psi} + \frac{\beta_\xi \xi}{\xi^*} \lambda e^{i \psi} \right] K_+ + \left[ \frac{\omega \alpha_\xi \xi^*}{\abs{\xi}} + \frac{\beta_\xi \xi^*}{\xi} \lambda e^{-i \psi} + (\beta_\xi + 1) \lambda e^{i \psi} \right] K_- \nonumber\\
        &&+ \frac{\xi^*}{\abs{\xi}} \alpha_\xi \lambda (K_{-}^{(b)} e^{-i \psi} + K_{-}^{(a)} e^{i \psi}) + \frac{\xi}{\abs{\xi}} \alpha_\xi \lambda (K_{+}^{(a)} e^{-i \psi} + K_{+}^{(b)} e^{i \psi}) + (2 \beta_\xi + 1) \lambda (J_+ e^{-i \psi} + J_- e^{i \psi}).
\end{eqnarray}
The coefficients of the operators $K_{\pm}$ will vanish if we set the values of the coherent state parameters $\tau$ and $\phi_\xi$ as
\begin{equation}
    \tau = \tanh^{-1}{\left( \frac{\lambda}{\omega} \right)}, \hspace{0.3cm} \phi_\xi = \psi.
\end{equation}
Then, we obtain the following reduced Hamiltonian
\begin{eqnarray}
        H' &=& 2[\omega \cosh{(\tau)} - \lambda \sinh{(\tau)}] K_0 + \lambda \cosh{(\tau)} (J_+ e^{-i \psi} + J_- e^{i \psi}) \nonumber\\
        &&- \lambda \sinh{(\tau)} (K_{+}^{(b)} + K_{-}^{(b)}) - \lambda \sinh{(\tau)} (K_{+}^{(a)} e^{-2i \psi} + K_{-}^{(a)} e^{2i \psi}).
\end{eqnarray}
Using that $\cosh{(\tau)} = \frac{\omega}{\sqrt{\omega^2 - \lambda^2}}$ and $\sinh{(\tau)} = \frac{\lambda}{\sqrt{\omega^2 - \lambda^2}}$ we get the expression
\begin{eqnarray}
        H' &=& 2 \sqrt{\omega^2 - \lambda^2} K_0 + \frac{\omega \lambda}{\sqrt{\omega^2 - \lambda^2}} (J_+ e^{-i \psi} + J_- e^{i \psi}) \nonumber\\
        &&- \frac{\lambda^2}{\sqrt{\omega^2 - \lambda^2}} (K_{+}^{(b)} + K_{-}^{(b)}) - \frac{\lambda^2}{\sqrt{\omega^2 - \lambda^2}} (K_{+}^{(a)} e^{-2i \psi} + K_{-}^{(a)} e^{2i \psi}).
\end{eqnarray}
Now, following a similar procedure, we can apply to this Hamiltonian a second tilting transformation based on the $SU(2)$ displacement operator defined as $D(\chi) = \exp(\chi J_+ - \chi^{*} J_-)$, with $\chi = - \rfrac{1}{2} \theta e^{-i \phi_\theta}$. Hence, with $H''$ and $\Psi''$ defined as $H''=D^{\dagger}(\chi)H'D(\chi)$ and $\Psi''=D^{\dagger}(\chi)\Psi'$ respectively, the similarity transformations of Eqs. (\ref{st3})-(\ref{st5}) of Appendix B let us obtain
\begin{equation}
    \begin{gathered}
        H'' = 2 \sqrt{\omega^2 - \lambda^2} K_0 \\
        + \frac{\omega \lambda}{\sqrt{\omega^2 - \lambda^2}} \left[ e^{-i \psi} \left[ - \frac{\chi^*}{\abs{\chi}} \alpha_\chi J_0 + \beta_\chi \left( J_+ + \frac{\chi^*}{\chi} J_- \right) + J_+ \right] + e^{i \psi} \left[ - \frac{\chi}{\abs{\chi}} \alpha_\chi J_0 + \beta_\chi \left( J_- + \frac{\chi}{\chi^*} J_+ \right) + J_- \right] \right] \\
        - \frac{\lambda^2}{\sqrt{\omega^2 - \lambda^2}} \left[ \left[ \frac{K_{-}^{(b)}}{2} \left( \cos{(2 \abs{\chi})} + 1 \right) - \frac{\chi^*}{2 \abs{\chi}} \sin{(2 \abs{\chi})} K_- - \frac{\chi^*}{2 \chi} \left( \cos{(2 \abs{\chi})} - 1 \right) K_{-}^{(a)} \right] \right. \\
        \left. + \left[ \frac{K_{+}^{(b)}}{2} \left( \cos{(2 \abs{\chi})} + 1 \right) - \frac{\chi}{2 \abs{\chi}} \sin{(2 \abs{\chi})} K_+ - \frac{\chi}{2 \chi^*} \left( \cos{(2 \abs{\chi})} - 1 \right) K_{+}^{(a)}  \right] \right] \\
        - \frac{\lambda^2}{\sqrt{\omega^2 - \lambda^2}} \left[ e^{-2i \psi} \left[ \frac{K_{+}^{(a)}}{2} \left( \cos{(2 \abs{\chi})} + 1 \right) + \frac{\chi^*}{2 \abs{\chi}} \sin{(2 \abs{\chi})} K_+ - \frac{\chi^*}{2 \chi} \left( \cos{(2 \abs{\chi})} - 1 \right) K_{+}^{(b)} \right] \right. \\
        \left. + e^{2i \psi} \left[ \frac{K_{-}^{(a)}}{2} \left( \cos{(2 \abs{\chi})} + 1 \right) + \frac{\chi}{2 \abs{\chi}} \sin{(2 \abs{\chi})} K_- - \frac{\chi}{2 \chi^*} \left( \cos{(2 \abs{\chi})} - 1 \right) K_{-}^{(b)} \right] \right],
    \end{gathered}
\end{equation}
where $\alpha_\chi = \sin{(2\abs{\chi})}$ and $\beta_\chi = \rfrac{1}{2} [ \cos{(2\abs{\chi})} - 1]$. \\
The coefficients of the operators $J_{\pm}$ will vanish if we consider that the values of the parameters are
\begin{equation}
    \theta = (2n + 1) \frac{\pi}{2} \hspace{0.3cm} \hbox{with} \hspace{0.3cm} n \in \boldsymbol{Z}, \hspace{0.3cm} \phi_\theta = \psi.
\end{equation}
Moreover, since $\cos{(\theta)} = 0$ and $\sin{(\theta)} = 1$, therefore we get the expression
\begin{equation}
    \begin{gathered}
        H'' = 2 \sqrt{\omega^2 - \lambda^2} K_0 + \frac{2 \omega \lambda}{\sqrt{\omega^2 - \lambda^2}} J_0 - \frac{\lambda^2}{\sqrt{\omega^2 - \lambda^2}} (K_{+}^{(b)} + K_{-}^{(b)}) - \frac{\lambda^2}{\sqrt{\omega^2 - \lambda^2}} (K_{+}^{(a)} e^{-2i \psi} + K_{-}^{(a)} e^{2i \psi}). \label{weak}
    \end{gathered}
\end{equation}

Now, if we consider the weak-coupling approximation $\lambda^2 \ll 1$ in Eq. (\ref{weak}), the Hamiltonian $H''$ is reduced to
\begin{equation}\label{H2}
    H'' = 2 \sqrt{\omega^2 - \lambda^2} K_0 + \frac{2 \omega \lambda}{\sqrt{\omega^2 - \lambda^2}} J_0.
\end{equation}
The operator $K_0$ represents the Hamiltonian of the two-dimensional harmonic oscillator and commutes with $J_0$. Therefore, the eigenfunctions of $H''$ are explicitly
\begin{equation}
    \Psi_{N, m} (r, \phi) = \frac{1}{\sqrt{\pi}} e^{im \phi} (-1)^{\frac{N - |m|}{2}} \sqrt{\frac{2 \left( \frac{N - |m|}{2} \right)!}{\left( \frac{N + |m|}{2} \right)!}} r^{|m|} L_{\frac{1}{2}(N - |m|)}^{|m|} (r^2) e^{- \frac{1}{2} r^2}, \label{sol1}
\end{equation}
or
\begin{equation}
    \Psi_{n_r, m} (r, \phi) = \frac{1}{\sqrt{\pi}} e^{im \phi} (-1)^{n_r} \sqrt{\frac{2 \left( n_r \right)!}{\left( n_r + |m| \right)!}} r^{|m|} L_{n_r}^{|m|} (r^2) e^{- \frac{1}{2} r^2}, \label{sol2}
\end{equation}
where $n_r = \frac{1}{2}(N - |m|)$ is the radial quantum number. Hence, the eigenfunctions of the Hamiltonian of two harmonic oscillators with identical free frequencies in the weak coupling approximation $\lambda^2 \ll 1$ are given by
\begin{equation}
\Psi=D(\xi)D(\chi)\Psi''=D(\xi)D(\chi) \Psi_{N, m} (r, \phi).
\end{equation}
It can be shown from the theory of irreducible representations of the $SU(1,1)$ and $SU(2)$ groups (see Appendix A) that the action of the operators $K_0$ and $J_0$ on the basis $|N,m\rangle$ are explicitly \cite{Nos1,Nos2}
\begin{equation}
K_0 \ket{N, m} = \frac{1}{2} (\hat{a}^{\dag} \hat{a} + \hat{b}^{\dag} \hat{b} + 1)\ket{N, m}=\frac{1}{2} (N + 1) \ket{N, m},\label{K0m}
\end{equation}
\begin{equation}
J_0|N,m\rangle=\frac{1}{2}\left(a^{\dag}a-b^{\dag}b\right)|N,m\rangle=\frac{m}{2}|N,m\rangle.\label{j0}
\end{equation}
Therefore, by substituting the Eqs. (\ref{K0m}) and (\ref{j0}) into Eq. (\ref{H2}) we obtain that the energy eigenvalues in the $\ket{N, m}$ basis for this problem are
\begin{equation}
    E = \sqrt{\omega^2 - \lambda^2} \left( N + 1 \right) + \frac{\omega \lambda}{\sqrt{\omega^2 - \lambda^2}} |m|.
\end{equation}
If we consider the case where $\lambda = 0$, this energy spectrum reduces to the expression
\begin{equation}
    E = \omega (N + 1) = \omega (2n_r + |m| + 1), \label{energy}
\end{equation}
which is the same energy spectrum of the two-dimensional harmonic oscillator in polar coordinates. Thus, we were able to obtain the eigenfunctions and the energy spectrum of the Hamiltonian of two isotropic oscillators with weak coupling by applying two tilting transformations. These transformations were based on the Schwinger bosonic realizations of the $su(1,1)$ and $su(2)$ Lie algebras.

\section{Mandel $Q$-parameter of the photon numbers $n_a$ and $n_b$ }

The statistical properties of light emission experiments are a fundamental piece to understand and describe the phenomena studied. Within these statistical tools is the Mandel $Q$-parameter \cite{Mandel,Mandel1}, which allows us to determine the nature of a weighting distribution. In particular, the Mandel parameter has been used to study photon statistics for different types of harmonic oscillators, as can be seen in Refs. \cite{Mahdifar,Amir,Amir2,Afshar,Dehghani}.

Thus, a simple way to gauge the nature of the photon statistics of any states is by computing the so-called $Q-$parameter, which is defined by \cite{Mandel,Mandel1}
\begin{equation}
    Q = \frac{\langle \hat{n}^{2} \rangle - \langle \hat{n} \rangle^{2}}{\langle \hat{n} \rangle} - 1, \label{Mandel}
\end{equation}
and where
\begin{equation}
    Q \left\{
    \begin{array}{lcc}
        >0, &   {\it super~Poissonian~distribution,}  &  \\
        \\ =0, &  {\it Poissonian~distribution~(coherent~state),}  &  \\
        \\ <0, &  {\it sub-Poissonian~distribution,}  & \\
        \\ =-1,&  {\it number~state.}
    \end{array}
    \right.
\end{equation}
Therefore, the Mandel $Q$-parameter allows us to measure the deviation from Poisson distribution, in order to distinguish quantum processes from classical ones.

In order to calculate the Mandel $Q$-parameter values, we shall take the states $|\Psi\rangle=D(\xi)D(\chi)|\Psi''\rangle$, where $|\Psi''\rangle=|N,m\rangle$. Hence, the Mandel parameter for the photon number $n_a$ is
\begin{equation}
    Q_{a}=\frac{\langle n^{2}_{a}\rangle-\langle n_{a}\rangle^{2}}{\langle n_{a}\rangle}-1,\label{Qa1}
\end{equation}
where
\begin{equation}
    \begin{gathered}\label{na2}
        \langle n_{a}^{2} \rangle = \langle (a^{\dag}a)^{2} \rangle = \langle \Psi|(a^{\dag}a)^{2} |\Psi \rangle = \langle \Psi''| \left[ D^{\dag}(\xi) D^{\dag}(\chi) a^{\dag}a D(\chi) D(\xi) \right]^{2} |\Psi'' \rangle,
    \end{gathered}
\end{equation}
\begin{equation}\label{n2a}
    \langle n_{a} \rangle^{2} = \langle a^{\dag}a \rangle^{2} = \left[ \langle \Psi''| D^{\dag}(\xi) D^{\dag}(\chi) a^{\dag}a D(\chi) D(\xi) |\Psi'' \rangle \right]^{2},
\end{equation}
and
\begin{equation}\label{na}
    \langle n_{a} \rangle = \langle a^{\dag}a \rangle = \langle \Psi''| D^{\dag}(\xi) D^{\dag}(\chi) a^{\dag}a D(\chi) D(\xi) |\Psi'' \rangle.
\end{equation}
Since $a^{\dag}a=K_{0} + J_{0} - \frac{1}{2}$, by using the $SU(1,1)$ displacement operator $D(\xi)$ and the $SU(2)$ displacement operator $D(\chi)$ we obtain that
\begin{equation}
    \begin{gathered}
        D^{\dag}(\chi) D^{\dag}(\xi) a^{\dag}a D(\xi) D(\chi) = D^{\dag}(\chi) D^{\dag}(\xi) \left[ K_{0} + J_{0} - \frac{1}{2} \right] D(\xi) D(\chi) = D^{\dag}(\chi) \left[ D^{\dag}(\xi) K_{0} D(\xi) + J_{0} - \frac{1}{2} \right] D(\chi) \\
        = \cosh(\tau) K_{0} - \frac{e^{-i \gamma}}{2} \sinh(\tau) \cos(\theta) K_{+} - \frac{e^{i\gamma}}{2} \sinh(\tau) \cos(\theta) K_{-} + \cos(\theta) J_{0} - \frac{e^{-i (\sigma + \gamma)}}{2} \sin(\theta) \sinh(\tau)  K^{(a)}_{+} \\
        + \frac{e^{i (\sigma - \gamma)}}{2} \sin(\theta) \sinh(\tau)  K^{(b)}_{+} - \frac{e^{i \sigma}}{2} \sin(\theta) J_{-} + \frac{e^{i (\gamma - \sigma)}}{2} \sin(\theta) \sinh(\tau)  K^{(b)}_{-} \\
        - \frac{e^{i (\sigma + \gamma)}}{2} \sin(\theta) \sinh(\tau)  K^{(a)}_{-} - \frac{e^{-i \sigma}}{2} \sin(\theta) J_{+} - \frac{1}{2},
    \end{gathered}
\end{equation}
where we have used the similarity transformations (\ref{st1})-(\ref{st4}) of Appendix B. Then, from equations (\ref{na2}), (\ref{n2a}) and (\ref{na}) we arrive to the following results
\begin{equation}
    \begin{gathered}
        \langle n^{2}_{a} \rangle = \cosh^{2}(\tau) \langle K^{2}_{0} \rangle_{N,m} + \cos^{2}(\theta) \langle J^{2}_0 \rangle_{N,m} + \cos(\theta) \cosh(\tau) \left[ \langle K_0J_0 \rangle_{N,m} + \langle J_{0}K_{0} \rangle_{N,m} \right] \\
        - \cosh(\tau) \langle K_0 \rangle - \cos(\theta) \langle J_0 \rangle + \frac{\sin^{2}(\theta)}{4} \left[ \langle J_{-}J_{+} \rangle_{N,m} + \langle J_{+}J_{-} \rangle_{N,m} \right] \\
        + \frac{\cos^{2}(\theta) \sinh^{2}(\tau)}{4} \left[ \langle K_{-}K_{+} \rangle_{N,m} + \langle K_{+}K_{-} \rangle_{N,m} \right] + \frac{\sin^{2}(\theta) \sinh^{2}(\tau)}{4} \left[ \langle K^{(a)}_{-}K^{(a)}_{+} \rangle_{N,m} + \langle K^{(a)}_{+}K^{(a)}_{-} \rangle_{N,m} \right] \\
        + \frac{\sin^{2}(\theta) \sinh^{2}(\tau)}{4} \left[ \langle K^{(b)}_{-}K^{(b)}_{+} \rangle_{N,m} + \langle K^{(b)}_{+}K^{(b)}_{-} \rangle_{N,m} \right] + \frac{1}{4},
    \end{gathered}
\end{equation}
while
\begin{equation}
    \begin{gathered}
        \langle n_{a} \rangle^{2} = \cosh^{2}(\tau) \langle K_{0} \rangle^{2}_{N,m} + \cos^{2}(\theta) \langle J_{0} \rangle^{2}_{N,m} + \cos(\theta) \cosh(\tau) \left[ \langle K_{0} \rangle_{N,m} \langle J_{0} \rangle_{N,m} + \langle J_{0} \rangle_{N,m} \langle K_{0} \rangle_{N,m} \right] \\
        - \cosh(\tau) \langle K_{0} \rangle_{N,m} - \cos(\theta) \langle J_{0} \rangle_{N,m} + \frac{1}{4},
    \end{gathered}
\end{equation}
and
\begin{equation}
    \langle n_{a} \rangle = \cosh(\tau) \langle K_{0} \rangle_{N,m} + \cos(\theta) \langle J_{0} \rangle_{N,m} - \frac{1}{2}.
\end{equation}
Therefore, the Mandel $Q_a$ parameter is given by
\begin{equation}
    \begin{gathered}
        Q_{a} = \frac{1}{4} \left[ \frac{\sin^{2}(\theta) \left[ \langle J_{-}J_{+} \rangle_{N,m} + \langle J_{+}J_{-} \rangle_{N,m} \right] + \cos^{2}(\theta) \sinh^{2}(\tau) \left[ \langle K_{-}K_{+} \rangle_{N,m} + \langle K_{+}K_{-} \rangle_{N,m} \right]}{\cosh(\tau) \langle K_{0} \rangle_{N,m} + \cos(\theta) \langle J_{0} \rangle_{N,m} - \frac{1}{2}} \right. \\
        \left. + \frac{\sin^{2}(\theta) \sinh^{2}(\tau) \left[ \langle K^{(a)}_{-}K^{(a)}_{+} \rangle_{N,m} + \langle K^{(a)}_{+}K^{(a)}_{-} \rangle_{N,m} \right] + \sin^{2}(\theta) \sinh^{2}(\tau) \left[ \langle K^{(b)}_{-}K^{(b)}_{+} \rangle_{N,m} + \langle K^{(b)}_{+}K^{(b)}_{-} \rangle_{N,m} \right]}{\cosh(\tau) \langle K_{0} \rangle_{N,m} + \cos(\theta) \langle J_{0} \rangle_{N,m} - \frac{1}{2}} \right] - 1.
    \end{gathered}
\end{equation}
By using the action of the $SU(1,1)$ and $SU(2)$ generators on the basis $|N,m\rangle$, we can compute the following expected values \cite{Nos1,Nos2}
\begin{equation}
    \begin{gathered}
        \langle K_{0} \rangle_{N,m} = \frac{N + 1}{2}, \quad\quad \langle J_{0} \rangle_{N,m} = \frac{m}{2}, \\
        \langle K_{+}K_{-} \rangle_{N,m} = \frac{N^{2} - m^{2}}{4}, \quad\quad \langle K_{-}K_{+} \rangle_{N,m} = \frac{N^{2} - m^{2}}{4} + N + 1, \\
        \langle J_{+}J_{-} \rangle_{N,m} = \frac{N^{2} - m^{2}}{4} + \frac{N + m}{2}, \quad\quad \langle J_{-}J_{+} \rangle_{N,m} = \frac{N^{2} - m^{2}}{4} + \frac{N - m}{2}, \\
        \langle K^{(a)}_{-}K^{(a)}_{+} \rangle_{N,m} = \frac{N^{2} + m^{2}}{16} + \frac{Nm}{8} + \frac{3}{8} (N + m) + \frac{1}{2}, \quad\quad \langle K^{(a)}_{+}K^{(a)}_{-} \rangle_{N,m} = \frac{N^{2} + m^{2}}{16} + \frac{Nm - N - m}{8}, \\
        \langle K^{(b)}_{-}K^{(b)}_{+} \rangle_{N,m} = \frac{N^{2} + m^{2}}{16} - \frac{Nm}{8} + \frac{3}{8} (N - m) + \frac{1}{2}, \quad\quad \langle K^{(b)}_{+}K^{(b)}_{-} \rangle_{N,m} = \frac{N^{2} + m^{2}}{16} - \frac{Nm + N - m}{8}, \label{12}
    \end{gathered}
\end{equation}
from which we find that the Mandel parameter $Q_a$ can be expressed as
\begin{equation}
    \begin{gathered}
        Q_{a} = \frac{1}{4} \left[ \frac{\sin^{2}(\theta) \left[ \frac{(N^2 - m^2)}{2} + N \right] + \cos^{2}(\theta) \sinh^{2}(\tau) \left[ \frac{(N^2 - m^2)}{2} + N + 1 \right] + \sin^{2}(\theta) \sinh^{2}(\tau) \left[ \frac{(N^2 + m^2)}{4} + \frac{N}{2} + 1 \right]}{\cosh(\tau) \frac{(N + 1)}{2} + \cos(\theta) \frac{m}{2} - \frac{1}{2}} \right] - 1.
    \end{gathered} \label{Q_a}
\end{equation}
Moreover, if we consider the results obtained in Sec. 2
\begin{equation}
    \begin{gathered}
        \cosh(\tau) = \frac{\omega}{\sqrt{\omega^{2} - \lambda^{2}}}, \quad\quad \sinh(\tau) = \frac{\lambda}{\sqrt{\omega^{2} - \lambda^{2}}} \quad\quad \cos(\theta) = 0, \quad\quad \sin(\theta) = 1, \label{21}
    \end{gathered}
\end{equation}
we obtain that the $Q_a$-parameter of Eq. (\ref{Q_a}) for the Hamiltonian of two isotropic oscillators with weak coupling can be written as
\begin{equation}
    \begin{gathered}\label{Qaf}
        Q_{a} = \frac{1}{8 \sqrt{\omega^2 - \lambda^2}} \left[ \frac{ \omega^2 \left[ 2(N^2 - m^2) + 4N \right] - \lambda^2 \left[ N(N + 2) - 3 m^2 - 4 \right]}{\omega (N + 1) - \sqrt{\omega^{2}-\lambda^{2}}} \right] - 1.
    \end{gathered}
\end{equation}
Following a procedure similar to that used in calculating the Mandel parameter $Q_a$, we can compute the Mandel parameter $Q_b$ for the photon number $n_b$ by considering that
\begin{equation}
b^{\dag}b = K_{0} - J_{0} - \frac{1}{2}.
\end{equation}
Therefore, by using the similarity transformations (\ref{st1})-(\ref{st4}) of Appendix B we obtain
\begin{equation}
     \begin{gathered}
        \langle n^{2}_{b} \rangle = \cosh^{2}(\tau) \langle K^{2}_{0} \rangle_{N,m} + \cos^{2}(\theta) \langle J^{2}_0 \rangle_{N,m} - \cos(\theta) \cosh(\tau) \left[ \langle K_0J_0 \rangle_{N,m} + \langle J_{0}K_{0} \rangle_{N,m} \right] \\
        - \cosh(\tau) \langle K_0 \rangle + \cos(\theta) \langle J_0 \rangle + \frac{\sin^{2}(\theta)}{4} \left[ \langle J_{-}J_{+} \rangle_{N,m} + \langle J_{+}J_{-} \rangle_{N,m} \right] \\
        + \frac{\cos^{2}(\theta) \sinh^{2}(\tau)}{4} \left[ \langle K_{-}K_{+} \rangle_{N,m} + \langle K_{+}K_{-} \rangle_{N,m} \right] + \frac{\sin^{2}(\theta) \sinh^{2}(\tau)}{4} \left[ \langle K^{(a)}_{-}K^{(a)}_{+} \rangle_{N,m} + \langle K^{(a)}_{+}K^{(a)}_{-} \rangle_{N,m} \right] \\
        + \frac{\sin^{2}(\theta) \sinh^{2}(\tau)}{4} \left[ \langle K^{(b)}_{-}K^{(b)}_{+} \rangle_{N,m} + \langle K^{(b)}_{+}K^{(b)}_{-} \rangle_{N,m} \right] + \frac{1}{4},
    \end{gathered}
\end{equation}
while
\begin{equation}
    \begin{gathered}
        \langle n_{b} \rangle^{2} = \cosh^{2}(\tau) \langle K_{0} \rangle^{2}_{N,m} + \cos^{2}(\theta) \langle J_{0} \rangle^{2}_{N,m} - \cos(\theta) \cosh(\tau) \left[ \langle K_{0} \rangle_{N,m} \langle J_{0} \rangle_{N,m} + \langle J_{0} \rangle_{N,m} \langle K_{0} \rangle_{N,m} \right] \\
        - \cosh(\tau) \langle K_{0} \rangle_{N,m} + \cos(\theta) \langle J_{0} \rangle_{N,m} + \frac{1}{4},
    \end{gathered}
\end{equation}
and
\begin{equation}
    \langle n_{b} \rangle = \cosh(\tau) \langle K_{0} \rangle_{N,m} - \cos(\theta) \langle J_{0} \rangle_{N,m} - \frac{1}{2}.
\end{equation}
Thus, the Mandel $Q_b$ parameter is given by
\begin{equation}
    \begin{gathered}
        Q_{b} = \frac{1}{4} \left[ \frac{\sin^{2}(\theta) \left[ \langle J_{-}J_{+} \rangle_{N,m} + \langle J_{+}J_{-} \rangle_{N,m} \right] + \cos^{2}(\theta) \sinh^{2}(\tau) \left[ \langle K_{-}K_{+} \rangle_{N,m} + \langle K_{+}K_{-} \rangle_{N,m} \right]}{\cosh(\tau) \langle K_{0} \rangle_{N,m} - \cos(\theta) \langle J_{0} \rangle_{N,m} - \frac{1}{2}} \right. \\
        \left. + \frac{\sin^{2}(\theta) \sinh^{2}(\tau) \left[ \langle K^{(a)}_{-}K^{(a)}_{+} \rangle_{N,m} + \langle K^{(a)}_{+}K^{(a)}_{-} \rangle_{N,m} \right] + \sin^{2}(\theta) \sinh^{2}(\tau) \left[ \langle K^{(b)}_{-}K^{(b)}_{+} \rangle_{N,m} + \langle K^{(b)}_{+}K^{(b)}_{-} \rangle_{N,m} \right]}{\cosh(\tau) \langle K_{0} \rangle_{N,m} - \cos(\theta) \langle J_{0} \rangle_{N,m} - \frac{1}{2}} \right] - 1.
    \end{gathered}
\end{equation}
Finally, by using the relations (\ref{12}) and (\ref{21}), the Mandel $Q_b$ results to be
\begin{equation}
    \begin{gathered}\label{Qbf}
        Q_{b} = \frac{1}{8 \sqrt{\omega^2 - \lambda^2}} \left[ \frac{ \omega^2 \left[ 2(N^2 - m^2) + 4N \right] - \lambda^2 \left[ N(N + 2) - 3 m^2 - 4 \right]}{\omega (N + 1) - \sqrt{\omega^{2} - \lambda^{2}}} \right] - 1.
    \end{gathered}
\end{equation}
Here it is important to note that $Q_a = Q_b$. The Mandel parameter remains an active field of study today, as can be seen in the Refs. \cite{Berrada,Jones,Nos5}. In particular, in Ref. {\cite{Nos5}}, it was computed the Mandel parameter of the non-degenerate parametric amplifier for the photon numbers $n_a$ and $n_b$. However, for that problem the parameters $Q_a$ and $Q_b$ turned out to be different.

Another relevant parameter in the study of photon statistics and which is closely related to the Mandel parameter is the second-order correlation function $g^2(0)$ \cite{Mandel,Mandel1,Glauber}. The second-order correlation function provides information on the bunching or the anti-bunching effects and is defined as
\begin{equation}
g^2(0)=\frac{\langle \hat{n}^{2} \rangle - \langle \hat{n} \rangle}{(\langle \hat{n} \rangle)^2}=\frac{Q}{\langle \hat{n} \rangle} + 1,
\end{equation}
where
\begin{equation}
    g^2(0) \left\{
    \begin{array}{lcc}
        >1, &   {\it bunching~effect,}  &  \\
        \\ <1, &  {\it anti-bunching~effect,}  &  \\
        \\ =1, &  {\it coherent~state.}
    \end{array}
    \right.
\end{equation}
From Eqs. (\ref{Qaf}) and (\ref{Qbf}) it can be shown that the second-order correlation function $g^2(0)$ for the photon numbers $n_a$ and $n_b$ are the same and it turns out to be
\begin{equation}\label{g20}
g_{a,b}^{2}(0) = \frac{Q_{a,b}}{\langle n_{a,b} \rangle} + 1 =  \frac{2 \omega^2 \left[ 3N (N+2) - m^2 + 8 \right] - \lambda^2 \left[ N (N+2) - 3m^2 + 8 \right] - 16 \omega (N+1) \sqrt{\omega^2 - \lambda^2}}{4 \left[ \omega (N+1) - \sqrt{\omega^2 - \lambda^2} \right]^2}.
\end{equation}
Here it is important to note that, derived from the conditions of Eq. (\ref{21}), for our problem $n_a=n_b$.

In Fig. 1 we have plotted the Mandel parameter $Q_a=Q_b$, given by equations (\ref{Qaf}) and (\ref{Qbf}) for $N=0,1,2,3,4,5,6$, $m=N,N-2,...,-N$ with the choice $\omega=4$ and $\lambda=0.5$. Note that the graph is cut by a horizontal plane. The states below this plane correspond to a sub-Poissonian distribution, while the states above it correspond to a super-Poissonian distribution. Therefore, both sub-Poissonian and super-Poissonian statistics are present for certain values of $N$ and $m$. Moreover, the ground state $N=0,m=0$ is approximately in a Poissonian distribution $(Q\approx0.00395263)$, while most the excited states for $N\geq4$, $-N<m<N$ are in the super-Poissonian distribution. The sub-Poissonian distribution is obtained for the excited states where $|m|=N$. This behavior remains practically the same for different values of $\omega$ and $\lambda$, as long as we are in the regime of weak coupling $\lambda\ll1$.

\begin{figure}[ht]
 \centering
    \includegraphics[width=0.95\textwidth]{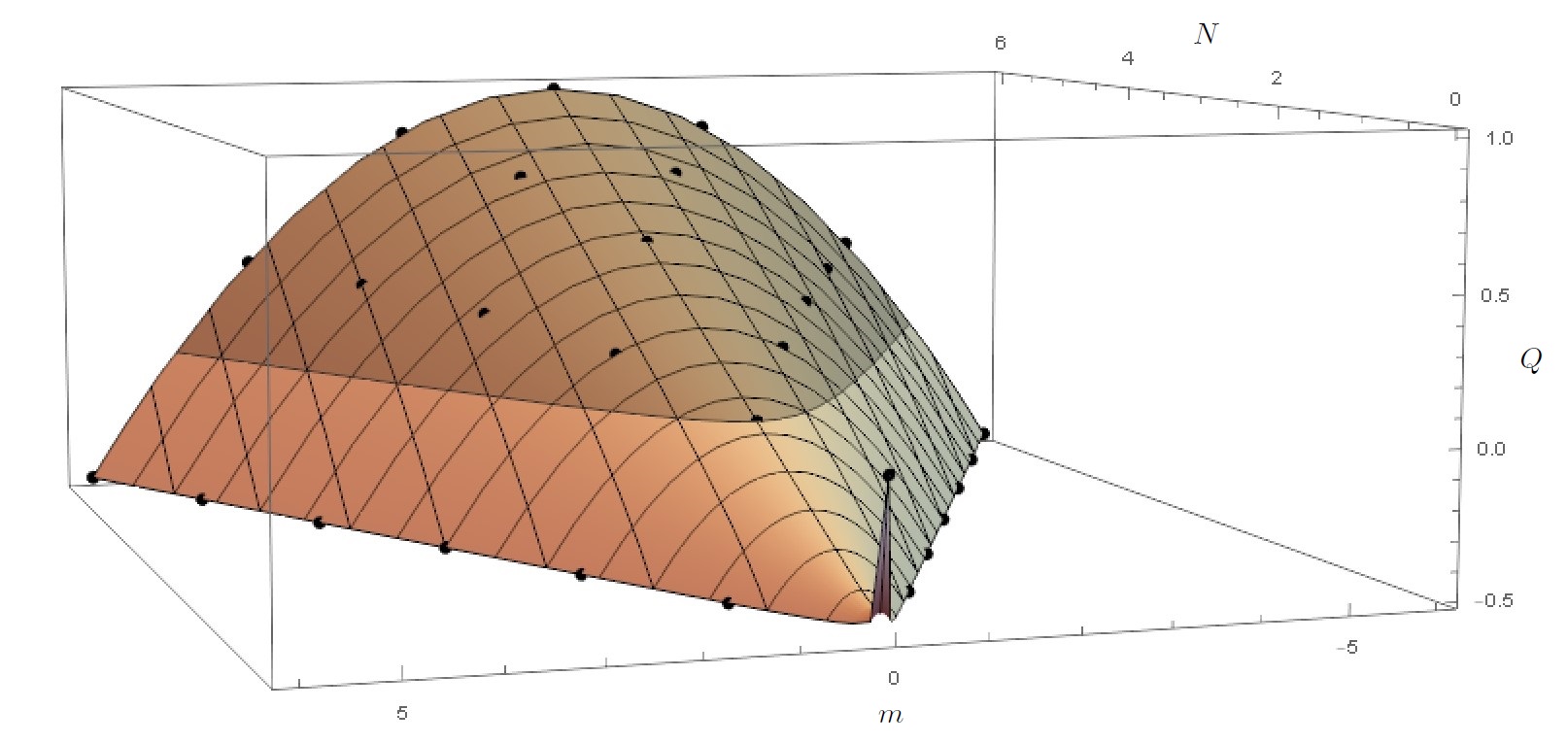}
  \caption{\footnotesize It is plotted the Mandel $Q$-parameter of two oscillators with weak coupling, for $N=0,1,2,3,4,5,6$, $m=N,N-2,...,-N$ with $\omega=4$, $\lambda=0.5$. The two regions separated by the horizontal plane represent the states where the system is in a super-Poissonian or sub-Poissonian distribution.}
\end{figure}

The second-order correlation function $g_{a,b}^2(0)$ of Eq. (\ref{g20}) is plotted in Fig. 2 under the same conditions of the Mandel parameter, that is $N=0,1,2,3,4,5,6$, $m=N,N-2,...,-N$ with the choice $\omega=4$ and $\lambda=0.5$. Here we observe that the problem of two oscillators within the weak coupling regime exhibits the bunching and anti-bunching effect. Moreover, the bunching effect occur for the states for which $-N<m<N$, whereas the anti-bunching effect is present for the states where $|m|=N$.

\begin{figure}[ht]
 \centering
    \includegraphics[width=0.95\textwidth]{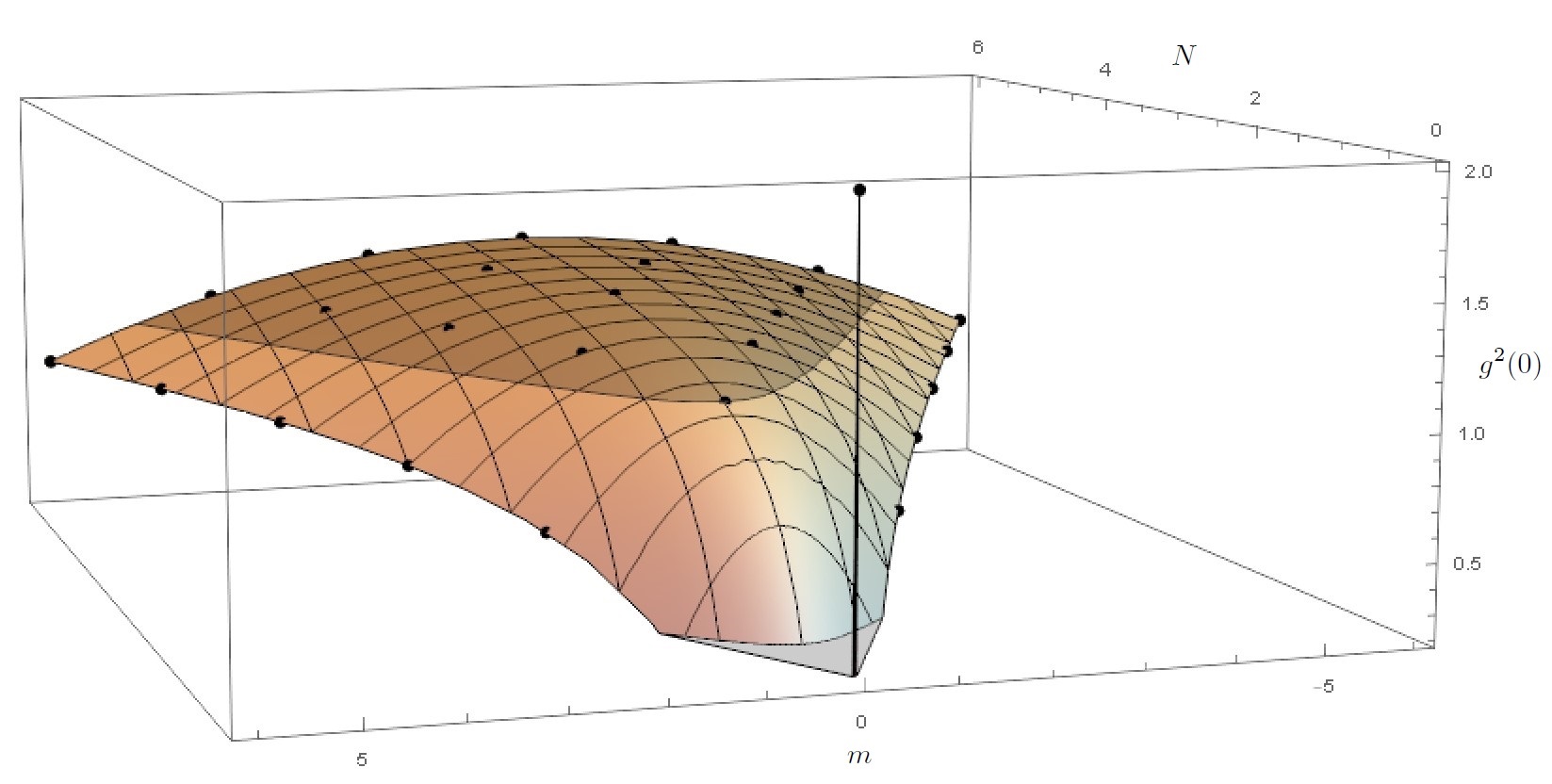}
  \caption{\footnotesize The second-order correlation function of two oscillators with weak coupling is represented for $N=0,1,2,3,4,5,6$, $m=N,N-2,...,-N$ with $\omega=4$, $\lambda=0.5$. The two regions represent the states for which the bunching and anti-bunching effects occur.}
\end{figure}

\section{Concluding remarks}

In this work we studied the Hamiltonian of two isotropic oscillators with weak coupling by using an $SU(1,1)\times SU(2)$ algebraic approach. In order to diagonalize our Hamiltonian, we applied two similarity transformations in terms of the $SU(1,1)$ and $SU(2)$ displacement operators. Assuming weak coupling between the oscillators $\lambda^2 \ll 1$, we were able to express the tilted Hamiltonian in terms of the Hamiltonian of the two-dimensional harmonic oscillator and the number difference operator $J_0$, which commute with each other. In this way, the energy spectrum and the eigenfunctions of our general Hamiltonian of two oscillators with weak coupling could be calculated.

Furthermore, we computed the similarity transformations of the number operators $n_a$ and $n_b$ in terms of the displacement operators of the $SU(1,1)$ and $SU(2)$ groups. This allowed us to obtain the Mandel parameters $Q_a$ and $Q_b$ and the second-order correlation function $g^2(0)$ of the photon numbers $n_a$ and $n_b$. We showed that for this problem, both Mandel parameters turned out to be the same $Q_a=Q_b$, and that most of the states of the system are in a super-Poissonian distribution. Only the states for which $|m|=n$ are in the sub-Poissonian distribution. In the case of the second-order correlation function it was showed that the bunching effect occurs for the states for which $-N<m<N$, whereas the anti-bunching effect is present for the states where $|m|=N$.

\section*{Acknowledgments}

This work was partially supported by SNII-M\'exico, EDI-IPN, SIP-IPN Project Number $20241764$. \\
We appreciate the comments made by the anonymous referees to improve our work.

\section*{Disclosures}

The authors declare no conflicts of interest.

\section*{Data Availability}

No data were generated or analyzed in the presented research.

\renewcommand{\theequation}{A.\arabic{equation}}
\setcounter{equation}{0}

\section*{Appendix A. The $SU(1,1)$ and $SU(2)$ group theory and its Perelomov coherent states}

The $su(1,1)$ and $su(2)$ Lie algebras satisfy the following commutation relations \cite{Vourdas}
\begin{eqnarray}
    [K_{0}, K_{\pm}] = \pm K_{\pm}, \quad\quad [K_{-}, K_{+}] = 2 K_{0}, \label{algebra1}
\end{eqnarray}
\begin{eqnarray}
    [J_{0}, J_{\pm}] = \pm J_{\pm}, \quad\quad [J_{+}, J_{-}] = 2 J_{0}. \label{algebra2}
\end{eqnarray}
In these expressions, the operators $K_{\pm}$, $K_0$ are the generators of the $su(1,1)$ Lie algebra, while the operators $J_\pm$ and $J_{0}$ are the generators of the $su(2)$ Lie algebra. The Casimir operators $K^2$ and $J^{2}$ for these algebras satisfy $[K^{2},K_{\pm}]=[K^{2},K_{0}]=0$ and $[J^{2},J_{\pm}]=[J^{2},J_{0}]=0$, and have the form
\begin{equation}
    K^2 = K_0^2 - \frac{1}{2} \left(K_+K_- + K_-K_+ \right), \quad\quad J^{2} = J_0^2 + \frac{1}{2} \left(J_+J_- + J_-J_+ \right).
\end{equation}
Now, the discrete representation of the $su(1,1)$ and $su(2)$ Lie algebra are given by
\begin{align}
    &K_{+} |k, n\rangle = \sqrt{(n + 1)(2k + n)} |k, n + 1 \rangle, \hspace{1.9cm} J_{+} |j, \mu \rangle = \sqrt{(j - \mu)(j + \mu + 1)} |j, \mu + 1 \rangle, \label{k+n}\\
    &K_{-} |k, n \rangle = \sqrt{n (2k + n - 1)} |k, n - 1 \rangle, \hspace{2.2cm} J_{-} |j, \mu \rangle = \sqrt{(j + \mu) (j - \mu + 1)} |j, \mu - 1 \rangle, \label{k-n}\\
    &K_{0} |k, n \rangle = (k + n) |k, n \rangle, \hspace{4.2cm} J_{0} |j, \mu \rangle = \mu |j, \mu \rangle, \label{k0n}\\
    &K^2 |k, n \rangle = k (k - 1) |k, n \rangle, \hspace{4cm} J^2 |j, \mu \rangle = j (j + 1) |j, \mu \rangle. \label{Cas}
\end{align}

The displacement operators $D(\xi)$ and $D(\chi)$ for these algebras are defined in terms of the creation and annihilation operators $\{K_+, K_- \}$ and $\{J_+,J_- \}$ as
\begin{equation}
    D(\xi)_{su(1,1)} = \exp(\xi K_{+} - \xi^{*} K_{-}), \quad\quad D(\chi)_{su(2)} = \exp(\chi J_{+} - \chi^{*} J_{-}), \label{do}
\end{equation}
where $\xi = - \rfrac{1}{2} \tau e^{-i \phi_\xi}$ and $\chi = - \rfrac{1}{2} \theta e^{-i \phi_\theta}$ with $- \infty < \tau, \theta < \infty$ and $0 \leq \phi_\xi, \phi_\chi \leq 2 \pi$. \\
Hence, the $SU(1,1)$ Perelomov number coherent states are defined as the action of the operator $D(\xi)$ onto an arbitrary state $|k, n \rangle$ as \cite{Nos1}
\begin{eqnarray}
    |\zeta_{\xi}, k, n \rangle & = & \sum_{s = 0}^\infty \frac{\zeta_{\xi}^s}{s!} \sum_{j = 0}^n \frac{(-\zeta_{\xi}^*)^j}{j!} e^{\eta_{\xi} (k + n - j)} \frac{\sqrt{\Gamma(2k + n) \Gamma(2k + n - j + s)}}{\Gamma(2k + n - j)} \nonumber\\
    &&\times \frac{\sqrt{\Gamma(n + 1) \Gamma(n - j + s + 1)}}{\Gamma(n - j + 1)} |k, n - j + s \rangle. \label{PNCS}
\end{eqnarray}
In an analogous way, the $SU(2)$ Perelomov number coherent states are defined as $D(\chi)|j ,\mu \rangle$ and explicitly are given by \cite{Nos2}
\begin{eqnarray}
    |\zeta_{\chi}, j ,\mu \rangle & = & \sum_{s = 0}^{j - \mu + n} \frac{\zeta_{\chi}^{s}}{s!} \sum_{n = 0}^{\mu + j} \frac{(-\zeta_{\chi}^*)^{n}}{n!} e^{\eta_{\chi} (\mu - n)} \frac{\Gamma(j - \mu + n + 1)}{\Gamma(j + \mu - n + 1)} \nonumber\\
    && \times \left[\frac{\Gamma(j + \mu + 1) \Gamma(j + \mu - n + s + 1)}{\Gamma(j - \mu + 1) \Gamma(j - \mu + n - s + 1)} \right]^{\frac{1}{2}} |j, \mu - n + s \rangle. \label{PNCS2}
\end{eqnarray}
where $\zeta_{\xi} = -\tanh( \frac{\tau}{2} ) e^{-i \phi_{\xi}}$, $\eta_{\xi} = \ln( 1 - |\zeta_{\xi}|^{2} )$ and $\zeta_{\chi} = -\tan( \frac{\theta}{2} ) e^{-i \phi_{\chi}}$ and $\eta_{\chi} = \ln( 1 + |\zeta_{\chi}|^{2} )$.

On the other hand, as it is well known, the bosonic annihilation $\hat{a}$, $\hat{b}$ and creation $\hat{a}^{\dag}$, $\hat{b}^{\dag}$ operators obey the commutation relations
\begin{equation}
    [\hat{a}, \hat{a}^{\dag}] = [\hat{b}, \hat{b}^{\dag}] = 1,
\end{equation}
\begin{equation}
    [\hat{a}, \hat{b}] = [\hat{a}^{\dag}, \hat{b}^{\dag}] = [\hat{a}^{\dag}, \hat{b}] = [\hat{a}, \hat{b}^{\dag}] = 0. \label{boson}
\end{equation}
These operators can be used appropriately to construct realizations of the $su(2)$ and $su(1,1)$ algebras. Thus, the $su(2)$ Lie algebra realization is given by the four operators \cite{Vourdas2}
\begin{equation}
    J_+ = \hat{a}^{\dag} \hat{b}, \quad\quad J_- = \hat{b}^{\dag} \hat{a}, \quad\quad J_0 = \frac{1}{2} (\hat{a}^{\dag} \hat{a} - \hat{b}^{\dag} \hat{b}), \quad\quad J^{2} = \frac{1}{4} N (N + 2), \label{su2}
\end{equation}
where $N = \hat{a}^{\dag} \hat{a} + \hat{b}^{\dag} \hat{b}$. \\
In terms of the $su(1,1)$ Lie algebra, we can obtain two different realizations \cite{Vourdas2}
\begin{equation}
    K_{+} = \hat{a}^{\dag} \hat{b}^{\dag}, \quad\quad K_{-} = \hat{b} \hat{a}, \quad\quad K_{0} = \frac{1}{2} (\hat{a}^{\dag} \hat{a} + \hat{b}^{\dag} \hat{b} + 1), \quad \quad K^{2} = J_{0}^{2} - \frac{1}{4}, \label{su11ab}
\end{equation}
and
\begin{equation}
    K_{+}^{(a)} = \frac{1}{2} \hat{a}^{\dag}{}^{2}, \quad\quad K_{-}^{(a)} = \frac{1}{2} \hat{a}^{2}, \quad\quad K_{0}^{(a)} = \frac{1}{2} \left( \hat{a}^{\dag} \hat{a} + \frac{1}{2} \right), \quad\quad K_{(a)}^{2} = -\frac{3}{16}. \label{su11a}
\end{equation}
Notice that for the second realization of the $su(1,1)$ Lie algebra, the Casimir operator $K_{(a)}^{2}$ is constant and the Bargmann index $k$ only can take the values $k=\frac{1}{4}$ and $k=\frac{3}{4}$.

\renewcommand{\theequation}{B.\arabic{equation}}
\setcounter{equation}{0}

\section*{Appendix B. Similarity transformations of the $SU(1,1)$ and $SU(2)$ group generators}

The $SU(1,1)$ and $SU(2)$ displacement operators defined in Appendix A and the Baker-Campbell-Hausdorff identity,
\begin{equation}
    e^{-A} B e^{A} = B + [B, A] + \frac{1}{2!} [[B, A], A] + \frac{1}{3!} [[[B, A], A], A] + ...,
\end{equation}
can be used to compute similarity transformations if they act on the $SU(1,1)$ and $SU(2)$ group generators. Thus, by using the $SU(1,1)$ displacement operator $D(\xi) = \exp(\xi K_+ - \xi^{*} K_-)$ and the commutation relations of Eqs. (\ref{algebra1}) and (\ref{algebra2}), we can compute the following similarity transformations for the $SU(1,1)$ operators $\{K_0,K_{\pm}\}$
\begin{equation}\label{st1}
    \begin{gathered}
        D^{\dagger}(\xi) K_0 D(\xi) = (2 \beta_\xi + 1) K_0 + \frac{\alpha_\xi \xi}{2 \abs{\xi}} K_+ + \frac{\alpha_\xi \xi^{*}}{2 \abs{\xi}} K_-, \\
        D^{\dagger}(\xi) K_+ D(\xi) = \frac{\xi^*}{\abs{\xi}} \alpha_\xi K_0 + \beta_\xi \left( K_+ + \frac{\xi^*}{\xi} K_- \right) + K_+, \\
        D^{\dagger}(\xi) K_- D(\xi) = \frac{\xi}{\abs{\xi}} \alpha_\xi K_0 + \beta_\xi \left( K_- + \frac{\xi}{\xi^*} K_+ \right) + K_-.
    \end{gathered}
\end{equation}
The similarity transformations for the $SU(2)$ group generators $\{J_0,J_{\pm}\}$ in terms of the $SU(1,1)$ displacement operator $D(\xi)$ are given by
\begin{equation} \label{st2}
    \begin{gathered}
        D^{\dagger}(\xi) J_0 D(\xi) = J_0, \\
        D^{\dagger}(\xi) J_+ D(\xi) = \frac{\xi^*}{\abs{\xi}} \alpha_\xi K_{-}^{(b)} + \frac{\xi}{\abs{\xi}} \alpha_\xi K_{+}^{(a)} + (2 \beta_\xi + 1) J_+, \\
        D^{\dagger}(\xi) J_- D(\xi) = \frac{\xi^*}{\abs{\xi}} \alpha_\xi K_{-}^{(a)} + \frac{\xi}{\abs{\xi}} \alpha_\xi K_{+}^{(b)} + (2 \beta_\xi + 1) J_-,
    \end{gathered}
\end{equation}
In these expressions $\alpha_\xi = \sinh{(2\abs{\xi})}$ and $\beta_\xi = \rfrac{1}{2} [ \cosh{(2\abs{\xi})} - 1]$.

Analogously, the $SU(2)$ displacement operator $D(\chi) = \exp(\chi J_+ - \chi^{*} J_-)$ can be used to obtain similarity transformations for the $SU(1,1)$ and $SU(2)$ Schwinger realizations of Eqs. (\ref{su2}), (\ref{su11ab}) and (\ref{su11a}). Therefore, for the $SU(2)$ Schwinger realization of Eq. (\ref{su2}) we obtain
\begin{equation}\label{st3}
    \begin{gathered}
        D^{\dagger}(\chi) J_0 D(\chi) = (2 \beta_\chi + 1) J_{0} + \frac{\chi}{2 |\chi|} \alpha_\chi J_{+} + \frac{\chi^{*}}{2 |\chi|} \alpha_\chi J_{-}, \\
        D^{\dagger}(\chi) J_+ D(\chi) = - \frac{\chi^*}{\abs{\chi}} \alpha_\chi J_0 + \beta_\chi \left( J_+ + \frac{\chi^*}{\chi} J_- \right) + J_+, \\
        D^{\dagger}(\chi) J_- D(\chi) = - \frac{\chi}{\abs{\chi}} \alpha_\chi J_0 + \beta_\chi \left( J_- + \frac{\chi}{\chi^*} J_+ \right) + J_-.
    \end{gathered}
\end{equation}
For the two-boson $SU(1,1)$ Schwinger realization (\ref{su11ab}) we can compute the following $SU(2)$ similarity transformations
\begin{equation}\label{st4}
    \begin{gathered}
        D^{\dagger}(\chi) K_0 D(\chi) = K_0, \\
        D^{\dagger}(\chi) K_+ D(\chi) = (2 \beta_\chi + 1) K_+ - \frac{\chi}{\abs{\chi}} \alpha_\chi K_+^{(a)} + \frac{\chi^{*}}{\abs{\chi}} \alpha_\chi K_+^{(b)}, \\
        D^{\dagger}(\chi) K_- D(\chi) = (2 \beta_\chi + 1) K_- - \frac{\chi^{*}}{\abs{\chi}} \alpha_\chi K_-^{(a)} + \frac{\chi}{\abs{\chi}} \alpha_\chi K_-^{(b)}.
    \end{gathered}
\end{equation}
Finally, we can obtain the $SU(2)$ similarity transformations for the one-boson $SU(1,1)$ Schwinger realization (\ref{su11a}). Therefore, if we assume that we have two realizations in terms of the $a$ and $b$ bosons
\begin{equation}\label{st5}
    \begin{gathered}
        D^{\dagger}(\chi) K_+^{(a)} D(\chi) = \left( \beta_\chi + 1 \right) K_{+}^{(a)} + \frac{\chi^*}{2 \abs{\chi}} \alpha_\chi K_+ - \frac{\chi^*}{\chi} \beta_\chi K_{+}^{(b)}, \\
        D^{\dagger}(\chi) K_-^{(a)} D(\chi) = \left( \beta_\chi + 1 \right) K_{-}^{(a)} + \frac{\chi}{2 \abs{\chi}} \alpha_\chi K_- - \frac{\chi}{\chi^*} \beta_\chi K_{-}^{(b)}, \\
        D^{\dagger}(\chi) K_+^{(b)} D(\chi) = \left( \beta_\chi + 1 \right) K_{+}^{(b)} - \frac{\chi}{2 \abs{\chi}} \alpha_\chi K_+ - \frac{\chi}{\chi^*} \beta_\chi K_{+}^{(a)}, \\
        D^{\dagger}(\chi) K_-^{(b)} D(\chi) = \left( \beta_\chi + 1 \right) K_{-}^{(b)} - \frac{\chi^*}{2 \abs{\chi}} \alpha_\chi K_- - \frac{\chi^*}{\chi} \beta_\chi K_{-}^{(a)}.
    \end{gathered}
\end{equation}
Here we have $\alpha_\chi = \sin{(2\abs{\chi})}$ and $\beta_\chi = \rfrac{1}{2} [ \cos{(2\abs{\chi})} - 1]$.


\begin{thebibliography} {99}

\bibitem{Bloch} S.C. Bloch, {\it Introduction to Classical and Quantum Harmonic Oscillators}, John Wiley \& Sons Inc., New York, 1997.

\bibitem{Fano} U. Fano, {\it Description of states in quantum mechanics by density matrix and operator techniques}, {\it Rev. Mod. Phys.} {\bf29}, 74 (1957).

\bibitem{Schweber} S.S. Schweber, {\it An Introduction to Relativistic Quantum Field Theory}, Row, Peterson and Company, Evanston, 1961.

\bibitem{Estes} L.E. Estes, T.H. Keil, and L.M. Narducci, {\it Quantum-Mechanical Description of Two Coupled Harmonic Oscillators}, {\it Phys. Rev.} {\bf175}, 286 (1968).

\bibitem{Fetter} A.L. Fetter, and J.D. Walecka, {\it Quantum Theory of Many Particle Systems}, McGraw-Hill, New York, 1971.

\bibitem{Kim} Y.S. Kim, {\it Observable Gauge Transformations in the Parton Picture}, {\it Phys. Rev. Lett.} {\bf63}, 348 (1989).

\bibitem{Han} D. Han, Y.S. Kim, and M.E. Noz, {\it Linear canonical transformations of coherent states in Wigner phase space. III. Two-mode states}, {\it Phys. Rev. A} {\bf41}, 6233 (1990).

\bibitem{Iachello} F. Iachello, and S. Oss, {\it Model of n Coupled Anharmonic Oscillators and Applications to Octahedral Molecules}, {\it Phys. Rev. Lett.} {\bf66}, 2976 (1991).

\bibitem{Han2} D. Han, Y.S. Kim, and M.E. Noz, {\it Illustrative example of Feynman's rest of the universe}, {\it Amer. J. Phys.} {\bf67}, 61 (1999).

\bibitem{Jakub} J.S. Prauzner-Bechcicki, {\it Two-mode squeezed vacuum state coupled to the common thermal reservoir}, {\it J. Phys. A: Math. Gen.} {\bf37}, L173 (2004).

\bibitem{Joshi} C. Joshi, A. Hutter, F.E. Zimmer, M. Jonson, E. Andersson, and P. \"Ohberg, {\it Quantum entanglement of nanocantilevers}, {\it Phys. Rev. A} {\bf82}, 043846 (2010).

\bibitem{Paz} J.P. Paz and A.J. Roncaglia, {\it Dynamics of the Entanglement Between Two Oscillators in the Same Environment}, {\it Phys. Rev. Lett.} {\bf100}, 220401 (2008).

\bibitem{Galve} F. Galve, L.A. Pachon, and D. Zueco, {\it Bringing Entanglement to the High Temperature Limit}, {\it Phys. Rev. Lett.} {\bf105}, 180501 (2010).

\bibitem{Fillaux} F. Fillaux, {\it Quantum entanglement and nonlocal proton transfer dynamics in dimers of formic acid and analogues}, {\it Chem. Phys. Lett.} {\bf408}, 302 (2005).

\bibitem{Romero} E. Romero et al., {\it Quantum coherence in photosynthesis for efficient solar-energy conversion}, {\it Nat. Phys.} {\bf10}, 676 (2014).

\bibitem{Fuller} F.D. Fuller et al., {\it Vibronic coherence in oxygenic photosynthesis}, {\it Nat. Chem.} {\bf6}, 706 (2014).

\bibitem{Halpin} A. Halpin et al., {\it Two-dimensional spectroscopy of a molecular dimer unveils the effects of vibronic coupling on exciton coherences}, {\it Nat. Chem.} {\bf6}, 196 (2014).

\bibitem{Ekert} A.K. Ekert, {\it Quantum Cryptography Based on Bell's Theorem}, {\it Phys. Rev. Lett.} {\bf67}, 661 (1991).

\bibitem{Bennett} C.H. Bennett and S.J. Wiesner, {\it Communication via One- and Two-Particle Operators on Einstein-Podolsky-Rosen States}, {\it Phys. Rev. Lett.} {\bf69}, 2881 (1992).

\bibitem{Shor} P.W. Shor, {\it Scheme for reducing decoherence in quantum computer memory}, {\it Phys. Rev. A} {\bf52}, R2493 (1995).

\bibitem{Samuel} L. Samuel, H. Braunstein, and J. Kimble, {\it Teleportation of Continuous Quantum Variables}, {\it Phys. Rev. Lett.} {\bf80}, 869 (1998).

\bibitem{Makarov} D.N. Makarov, {\it Coupled harmonic oscillators and their quantum entanglement}, {\it Phys. Rev. E} {\bf97}, 042203 (2018).

\bibitem{Gerryberry} C.C. Gerry, {\it Berry's phase in the degenerate parametric amplifier}, {\it Phys. Rev. A} {\bf39}, 3204 (1989).

\bibitem{Nos1} D. Ojeda-Guill\'en, R.D. Mota, and V.D. Granados, {\it The $SU(1,1)$ Perelomov number coherent states and the non-degenerate parametric amplifier}, {\it J. Math. Phys.} {\bf55}, 042109 (2014).

\bibitem{Nos2} D. Ojeda-Guill\'en, R.D. Mota, and V.D. Granados, {\it Spin number coherent states and the problem of two coupled oscillators}, {\it Commun. Theor. Phys.} {\bf64}, 34 (2015).

\bibitem{Nos3} E. Chore\~no, and D. Ojeda-Guill\'en, {\it $Sp(4,R)$ algebraic approach of the most general Hamiltonian of a two-level system in two-dimensional geometry}, {\it Eur. Phys. J. Plus} {\bf134}, 606 (2019).

\bibitem{Nos4} E. Chore\~no, R. Valencia, and D. Ojeda-Guill\'en, {\it Algebraic approach and Berry phaseof a Hamiltonian with a general $SU(1,1)$ symmetry}, {\it J. Math. Phys.} {\bf62}, 071701 (2021).

\bibitem{Mandel} L. Mandel, and E. Wolf, {\it Optical Coherence and Quantum Optics}, Cambridge Univ. Press, Cambridge, 1995.

\bibitem{Mandel1} L. Mandel, {\it Sub-Poissonian photon statistics in resonance fluorescence}, {\it Opt. Lett.} {\bf4}, 205 (1979).

\bibitem{Mahdifar} A. Mahdifar, R. Roknizadeh, and M.H. Naderi, {\it Geometric approach to nonlinear coherent states using the Higgs model for harmonic oscillator}, {\it J. Phys. A: Math. Gen.} {\bf39}, 7003 (2006).

\bibitem{Amir} N. Amir, and S. Iqbal, {\it Coherent states for nonlinear harmonic oscillator and some of its properties}, {\it J. Math. Phys.} {\bf56}, 062108 (2015).

\bibitem{Amir2} N. Amir, and S. Iqbal, {\it Barut-Girardello Coherent States for Nonlinear Oscillator with Position-Dependent Mass}, {\it Commun. Theor. Phys.} {\bf66}, 41 (2016).

\bibitem{Afshar} D. Afshar, A. Motamedinasab, A. Anbaraki, and M. Jafarpour, {\it Even and odd coherent states of supersymmetric harmonic oscillators and their nonclassical properties}, {\it Int. J. Mod. Phys. B} {\bf30}, 1650026 (2016).

\bibitem{Dehghani} A. Dehghani, B. Mojaveri, and A.A. Alenabi, {\it Excitation and depression of coherent state of the simple harmonic oscillator}, {\it J. Math. Phys.} {\bf60}, 083501 (2019).

\bibitem{Berrada} M. Algarni, K. Berrada, S. Abdel-Khalek, and H. Eleuch, {\it Parity deformed tavis-cummings model: Entanglement, parameter estimation and statistical properties}, {\it Mathematics} {\bf10}, 3051 (2022).

\bibitem{Jones} C. Jones, J. Xavier, S.V. Kashanian, M. Nguyen, I. Aharonovich, and F. Vollmer, {\it Time-Dependent Mandel Q Parameter Analysis for a Hexagonal Boron Nitride Single Photon Source}, {\it Opt. Express} {\bf31}, (2023).

\bibitem{Nos5} J.C. Vega, E. Chore\~no, D. Ojeda-Guill\'en, and R.D. Mota, {\it Berry phase and the Mandel parameter of the non-degenerate parametric amplifier}, {\it JOSA B} {\bf41}, 1084 (2024).

\bibitem{Glauber} R.J. Glauber, {\it Quantum Theory of Optical Coherence: Selected Papers and Lectures}, John Wiley \& Sons, Weinheim, 2007.

\bibitem{Vourdas} A. Vourdas, {\it $SU(2)$ and $SU(1,1)$ phase states}, {\it Phys. Rev. A} {\bf41}, 1653 (1990).

\bibitem{Vourdas2} A. Vourdas, {\it Analytic representations in quantum mechanics}, {\it J. Phys. A: Math. Gen.} {\bf39}, R65 (2006).



\end{thebibliography}
\end{document}